\documentclass[a4paper,11pt]{article}
\pdfoutput=1 
\usepackage{jheppub}

\preprint{}

\usepackage{amsmath}
\usepackage{amsfonts}  
\usepackage{amssymb}
\usepackage{slashed}
\usepackage{stmaryrd}
\usepackage{bm}
\usepackage{graphicx}%
\usepackage[T1]{fontenc} 
\usepackage{mathrsfs}
\usepackage{yfonts}
\usepackage{pifont}

\DeclareMathAlphabet{\mathpzc}{OT1}{pzc}{m}{it}

\usepackage{graphicx}
\usepackage{dcolumn}
\usepackage{epsfig}

\newcommand{\beq}{\begin{equation}} 
\newcommand{\eeq}{\end{equation}} 
\newcommand{\bega}{\begin{eqnarray}} 
\newcommand{\ega}{\end{eqnarray}} 

\newcommand{\dhd}{{\textstyle d}
\lower.03ex\hbox{\kern-0.38em$^{\scriptstyle-}$}\kern-0.05em{}}
\newcommand{\dbar}{{\textstyle \delta}
\lower.03ex\hbox{\kern-0.38em$^{\scriptstyle-}$}\kern-0.05em{}}
\newcommand{\half}{{1\over 2}}

\newcommand{\cald}{{\cal D}}

\newcommand{\hacalo}{{\hat {\cal O}}}

\newcommand{\tilD}{{\tilde D}}

\newcommand{\tigma}{\tilde {\sigma}}

\newcommand{\pizf}{\mathpzc{F}}
\newcommand{\pizg}{\mathpzc{G}}

\newcommand{\ve}{\varepsilon}

\newcommand{\vigma}{\varsigma}

\pdfoutput=1
\abstract{Within the framework of rapidity-only small-$x$ TMD factorization, the evolution with respect to the rapidity cutoff is initially governed by Sudakov double logarithms and subsequently by BFKL/BK single logarithms. The evolution equation proposed in this paper correctly reproduces both the Sudakov and BFKL limits, while providing a consistent interpolation between these two regimes.
 }
\arxivnumber{}

\emailAdd{balitsky@jlab.org}

\preprint{}

\title{Interpolation between Sudakov and BFKL rapidity evolutions for TMD factorization at small $x$}

\author{ Ian Balitsky}

\affiliation{%
Physics Dept., Old Dominion University, Norfolk, VA 23529 
}

\begin{document}
\maketitle

\flushbottom

\section{Introduction\label{aba:sec1}}

The production of particles in hadron--hadron collisions in the kinematic regime where the transverse momentum of the observed final state is much smaller than its invariant mass is described by transverse-momentum-dependent (TMD) factorization. Prominent examples include the Drell--Yan process and Higgs production via gluon fusion. The factorization formula reads \cite{Collins:1984kg,Ji:2004wu,Collins:2011zzd, Collins:2014jpa}
\begin{eqnarray}
&&\hspace{-1mm}
{d\sigma\over d\eta d^2q_\perp}~=~
\sum_f\!\int\! d^2b_\perp e^{i(q,b)_\perp}
\cald_{f/A}(x_A,b_\perp,\eta_a)\cald_{f/B}(x_B,b_\perp,\eta_b)
\nonumber\\
&&\hspace{1mm}
\times~
\sigma_{ff\rightarrow X}(\eta, \eta_a,\eta_b)
~+~{\rm power~corrections}~+~{\rm Y\text{-}terms},
\label{TMDf}
\end{eqnarray}
where $\eta=\frac{1}{2}\ln(q^+/q^-)$ is the rapidity, $\cald_{f/h}(x,b_\perp,\eta_i)$ denotes the TMD parton distribution of parton $f$ in hadron $h$ with rapidity cutoff $\eta_i$, and $\sigma_{ff\rightarrow X}$ is the partonic cross section for producing a state $X$ with invariant mass $Q^2=q^2\gg q_\perp^2$.

In the Collins--Soper--Sterman (CSS) framework \cite{Collins:1984kg,Collins:2011zzd}, TMD distributions are regularized by ultraviolet and rapidity cutoffs, and the corresponding Sudakov logarithms are obtained through evolution in the scales $\mu_{\rm UV}$ and $\eta_i$. At small Bjorken $x_{A,B}$, however, this approach becomes inadequate and must be replaced by a rapidity-factorization framework appropriate for the small-$x$ regime.

Rapidity-only TMD factorization was introduced in \cite{Balitsky:2017flc} and further developed in \cite{Balitsky:2017gis,Balitsky:2019ayf,Balitsky:2020jzt,Balitsky:2021fer,Balitsky:2022vnb,Balitsky:2023hmh}. To illustrate the method, consider Higgs production via gluon fusion at LHC energies with $x_A\sim x_B\sim 0.01$. As in Ref.~\cite{Balitsky:2017flc}, we approximate the gluon--gluon--Higgs triangle by an effective local vertex.

The key idea of rapidity-only factorization is to decompose quark and gluon fields into three sectors: projectile fields with small components along $p_B$, target fields with small components along $p_A$, and central fields. TMD factorization is obtained by integrating over the central fields in the background of frozen projectile and target fields.

At tree level, this procedure yields the classical solution of the Yang--Mills equations with sources provided by the projectile and target fields. The general solution, corresponding to the scattering of two color-glass condensates, is not known in closed form. Nevertheless, in the limit $q_\perp^2/Q^2\ll1$, one can construct a systematic expansion in this parameter, generating a series of TMD operators suppressed by powers of $q_\perp^2/Q^2$ that produce power corrections to eq.~(\ref{TMDf}).
\footnote{In our example $Q=m_H$=125 GeV is the Higgs mass, but we keep the notation $Q$ since the approach can be applied to other cases, 
such as production of $\bar{b}b$ pair with $Q\sim 10$ GeV.}

\begin{figure}[htb]
\begin{center}
\includegraphics[width=111mm]{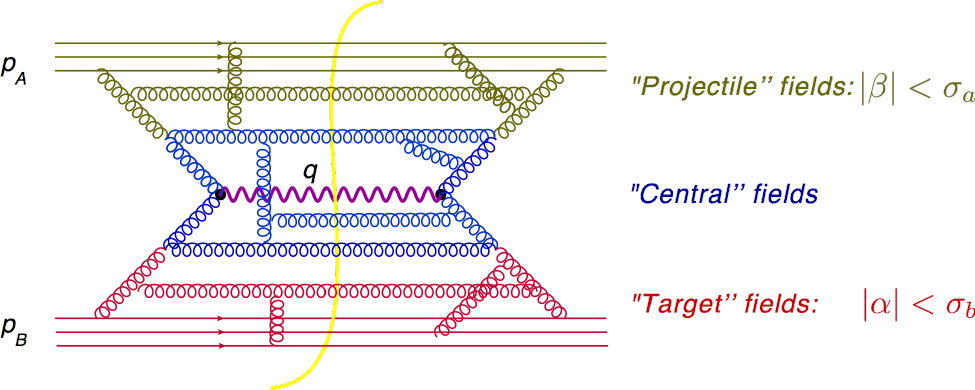}
\end{center}
\caption{Rapidity factorization for a typical diagram for Higgs production by gluon fusion. Gluon-gluon-Higgs triangle is approximated by a local vertex. \label{fig:real1}}
\end{figure}

Beyond tree level, the implementation of rapidity cutoffs and the associated evolution becomes essential. Using the Sudakov 
decomposition\footnote{Strictly speaking, $p = \alpha p_1 + \beta p_2 + p_\perp$ where $p_1$ and $p_2$ are light-like vectors close to $p_A$ and $p_B$.}
$p = \alpha p_A + \beta p_B + p_\perp$,
one defines projectile gluons by $\beta<\sigma_a$ and target gluons by $\alpha<\sigma_b$, see Fig. \ref{fig:real1}. 
Functional integration over projectile (target) fields yields $\cald_{f/A}(x_A,b_\perp,\ln\sigma_a)$ ($\cald_{f/B}(x_B,b_\perp,\ln\sigma_b)$), while integration over central fields produces the coefficient function $\sigma_{ff\rightarrow X}(\ln\sigma_a,\ln\sigma_b)$. The result of calculation of the 
integral over central fields for hadronic tensor has the form 
\begin{eqnarray}
&&\hspace{-2mm}
W(q)~=~\int\! db_\perp~e^{i(q,b)_\perp}
W( x_A, x_B,b_\perp),
\nonumber\\
&&\hspace{-1mm}
W( x_A, x_B,b_\perp)
~=~{\pi^2\over 2}Q^2\pizg_{ij}^{\sigma_a}( x_A,b_\perp;p_A)\pizg^{ij;\sigma_b}( x_B,b_\perp;p_B)
\nonumber\\
&&\hspace{36mm}
\times~\exp\Big\{{\alpha_sN_c\over 2\pi}\Big[
\ln^2{b_\perp^2s\sigma_a\sigma_b\over 4}
-2\big(\ln{ x_A\over\sigma_b}+\gamma\big)\big(\ln{ x_B\over\sigma_a}+\gamma\big)+{\pi^2\over 2}\Big]\Big\}
\nonumber\\
&&\hspace{36mm}
+~{\rm NLO~ terms}\sim O\big(\alpha_s^2)~+~{\rm power~corrections}~~
\label{momentumresult}
\end{eqnarray}
where $\pizg_{ij}^{\sigma_a},\pizg_{ij}^{\sigma_b}$ 
are related to gluon TMD $D_g^{\sigma_a}(x,b_\perp)$  by Eq. (\ref{operde}) below.

A natural strategy is to maximize the phase space attributed to central fields, leaving TMD matrix elements defined at small cutoffs $\sigma_a$ and $\sigma_b$ free of large logarithms. In this case, all large logarithms are resummed into the coefficient function, after which nonperturbative input for TMDs may be supplied, e.g. from models or lattice calculations.

This program, however, is not compatible with the small-$x$ regime. The factorization formula (\ref{momentumresult}) is valid for 
$\sigma_a \geq \tigma_p = {4b_\perp^{-2}\over x_A s}$ and $\sigma_b \geq \tigma_t = {4b_\perp^{-2}\over x_B s}$, where it correctly reproduces Sudakov double logarithms. However, the TMD distributions at these cutoffs contain large BFKL logarithms, $\ln(\sigma_a/\sigma_0)=\ln x_A$ and $\ln(\sigma_b/\sigma_0)=\ln x_B$, with $\sigma_0 = q_\perp^2/s$ the endpoint of BFKL \cite{Fadin:1975cb,Balitsky:1978ic}
or BK \cite{Balitsky:1995ub,Kovchegov:1999yj,Kovchegov:1999ua} evolution. Since BK evolution is nonlinear and involves nontrivial transverse dynamics, it cannot be absorbed into eq.~(\ref{TMDf}).

The appropriate procedure is therefore to apply eq.~(\ref{TMDf}) down to $\sigma_a$ and $\sigma_b$, and subsequently 
evolve $\cald^{\sigma_a}(u,b_\perp)$ and $\cald^{\sigma_b}(u,b_\perp)$ using BFKL/BK evolution down to $\sigma_0$. At this scale, the gluon density is typically modeled (e.g. via MV \cite{McLerran:1993ni,McLerran:1993ka} or geometrical scaling \cite{Golec-Biernat:1998zce} models), yielding the final cross section. 
Such approach - convolution of two different evolutions - goes under the name “hybrid factorization” \cite{Dumitru:2005gt,Chirilli:2011km,Chirilli:2012jd,Mueller:2013wwa}.
The drawback of this “hybrid factorization” is the dependence on the matching scales $\sigma_a$ and $\sigma_b$.

Ideally, one would employ a single evolution equation interpolating between Sudakov and BK dynamics. Such an equation was derived in \cite{Balitsky:2015qba}, reproducing the DGLAP, Sudakov, and BK limits, but its complexity renders practical solutions difficult. For the present small-$x$ analysis, the exact DGLAP limit is not essential, and an interpolation between Sudakov and BK evolution suffices. In the linearized (BFKL) approximation, this interpolation can be expressed in a relatively simple form, which will be presented in the next section.

 \section{Evolution equation for gluon TMD}
Gluon TMD operators are defined as follows
\begin{eqnarray}
&&\hspace{-2mm}
\hacalo_{ij}^{\sigma_a}(z_2^+,z_{2_\perp};z_1^+,z_{1_\perp})
~=~\pizf^a_i(z_2)
[z_2-\infty^+,z_1-\infty^+]^{ab}\pizf^b_j(z_1)\Big|_{z_2^-=z_1^-=0}\,,
\label{kalo}
\\
&&\hspace{-2mm}
\hacalo_{ij}^{\sigma_b}(z_2^-,z_{2_\perp};z_1^-,z_{1_\perp})
~=~
\pizf^a_i(z_2)
[z_2-\infty^-,z_1-\infty^-]^{ab}\pizf^b_j(z_1)\Big|_{z_2^+=z_1^+=0}\,,
\nonumber
\end{eqnarray}
where
\begin{eqnarray}
&&\hspace{-1mm}
\pizf^{i,a}(z_\perp,z^+)~\equiv~gF^{- i,m}(z)[z^+,-\infty^+]_z^{ma}\Big|_{z^-=0},~~
\nonumber\\
&&\hspace{-2mm}
\pizf^{i,a}(z_\perp,z^-)~\equiv~gF^{+i,m}(z)[z^-,-\infty^-]^{ma}\Big|_{z^+=0}
\nonumber
\end{eqnarray}
are gluon stress tensors with attached Wilson lines going to $-\infty$ in ``+’’ and ``-’’ directions. 
Here $[x,y]$ denotes straight-line gauge link connecting points $x$ and $y$.  In the framework of rapidity-only factorization these TMDs operators are defined with 
cutoffs $\sigma_a$ and $\sigma_b$. Roughly speaking,  gluons emitted from $\hacalo_{ij}^{\sigma_a}$ have $|\alpha|<\sigma_a$ and correspondingly gluons emitted from 
 $\hacalo_{ij}^{\sigma_b}$ have $|\beta|<\sigma_b$. More accurately, these cutoffs are defined by ``point-splitting’’ regularization discussed in Ref. \cite{Balitsky:2023hmh}.

 Gluon TMDs   \cite{Mulders:2000sh}  are defined as matrix elements  of these operators
\begin{eqnarray}
&&\hspace{-1mm}
{2\over s}\langle p_A|\hacalo_{ij}^{\sigma_a}( x_A,b_\perp) |p_A\rangle~
\equiv~\sqrt{2\over s}\!\int\! dz^-e^{i x_A\sqrt{s\over 2} z^-}\langle p_A|\hacalo_{ij}^{\sigma_a}(z^-,0^-,b_\perp) |p_A\rangle
\nonumber\\
&&\hspace{35mm}
=~-\pi g^2x_A\pizg_{ij}^{\sigma_a}(x_A,b_\perp),
\nonumber\\
&&\hspace{-1mm}
\pizg_{ij}^{\sigma_a}(\alpha,b_\perp)~=~g_{ij}D^{\sigma_a}_g(\alpha,b_\perp)
+{1\over 2m_N^2}(2\partial_i\partial_j+g_{ij}\partial_\perp^2)H_g^{\sigma_a}(\alpha,b_\perp)
\label{operde}
\end{eqnarray}
and similarly for the target TMDs.

In the Sudakov region $\sigma_a\gg\tigma_p$ the projectile gluon TMD satisfies the equation \cite{Balitsky:2022vnb}
\begin{eqnarray}
&&\hspace{-1mm}
\sigma{d\over d\sigma}D_g^\sigma(\alpha,z_\perp)
~=~-a_s\big[\ln \sigma_a\alpha s{z_\perp^2\over 4}+\gamma_E\big]D_g^\sigma(\alpha,z_\perp)
\label{sudevol}
\end{eqnarray}
Hereafter we use the notation $a_s\equiv {\alpha_sN_c\over\pi}$.

On the other hand, at $\sigma_a\ll\tigma_p$ this TMD satisfies the BFKL equation
\footnote{As demonstrated in Refs. \cite{Dominguez:2011wm,Balitsky:2015qba,Balitsky:2016dgz} gluon TMD is evolving non-linearly according to the BK equation,  but
here for simplicity we consider linearized version of BK which is BFKL.}
\begin{eqnarray}
&&\hspace{-1mm}
\sigma{d\over d\sigma}D_g^\sigma(\alpha,z_\perp)~=~{a_s\over\pi}\!\int\!dz\Big[{1\over (z-z’)^2}D_g^\sigma(\alpha,z’_\perp)
-{(z,z’)\over (z-z’)^2{z’}^2}D_g^\sigma(\alpha,z_\perp)\Big]
\label{bfklevol}
\end{eqnarray}
In the rest of this Section all vectors will be 2-dim Euclidean so hereafter we omit sign $\perp$ from vectors. Also, $(x,y)\equiv x\cdot y>0$.

 One can interpolate between these two evolutions using the equation
\beq
\sigma{d\over d\sigma}D_g^\sigma(\alpha,z)~=~{a_s\over\pi}\!\int\!dz\Big[{e^{-\alpha\sigma {s\over 4} (z-z’)^2}\over (z-z’)_\perp^2}D_g^\sigma(\alpha,z’)
-{(z,z’)\over (z-z’)^2{z’}^2}D_g(\alpha,z)\Big]
\label{interpol}
\eeq
Obviously, at $\sigma\ll \sigma_a = {4z^{-2}\over x_A s}$ this equation reduces to the BFKL equation (\ref{bfklevol}). On the other hand, if 
$\sigma\gg \sigma_a = {4z^{-2}\over x_A s}$ one obtains
\begin{eqnarray}
&&\hspace{-1mm}
\sigma{d\over d\sigma}D_g^\sigma(\alpha,z)~\simeq~{a_s\over\pi}D_g^\sigma(\alpha,z)
\!\int\!dz\Big[{e^{-\alpha\sigma {s\over 4} (z-z’)^2}\over (z-z’)^2}-{(b,z)\over (z-z’)^2{z’}^2}\Big]
\nonumber\\ 
&&\hspace{27mm}
~=~-a_s\big[\ln \sigma_a\alpha s{z^2\over 4}+\gamma_E\big]D_g^\sigma(\alpha,z)
\label{evold}
\end{eqnarray}
which is Eq. (\ref{sudevol}).
The hope is to solve Eq. (\ref{interpol}),  at least numerically, and get proper matching between Sudakov and BFKL evolutions. 
For now, one can calculate corrections coming from interpolation for these two cases.

\subsection{Correction to Sudakov evolution}
To estimate the corrections we will use the MV model for WW gluon TMDs \cite{Dominguez:2011wm}
\begin{eqnarray}
&&\hspace{-1mm}
\alpha D_g^\sigma(\alpha,z)
~=~{C\over z^2}\Big[1-\exp\Big(-{Q_s^2z^2\over 4}\Big)\Big]
\label{mvodel}
\end{eqnarray}
with $Q_s\sim 0.6$GeV  at $x\sim 0.0$1 as suggested in \cite{Boer:2016fqd}. The value of the constant $C$ is irrelevant for our purposes.

In the Sudakov region $\sigma_a\gg\tigma_p = {4b^{-2}\over x_A s}$ one can rewrite Eq. (\ref{interpol}) as 
\begin{eqnarray}
&&\hspace{-1mm}
\sigma{d\over d\sigma}D_g^\sigma(\alpha,b;)
~=~-a_s\big[\ln\alpha\sigma {s\over 4} b^2+\gamma_E\big]D_g^\sigma(\alpha,b)
\nonumber\\ 
&&\hspace{-1mm}
+~{a_s\over\pi}\!\int\!dz {e^{-\alpha\sigma {s\over 4} z^2}\over z^2}\big[D_g^\sigma(\alpha,z+b)-D_g^\sigma(\alpha,b;)\big]
\nonumber\\ 
&&\hspace{-1mm}
\simeq~-a_s\big[\ln \alpha\sigma {s\over 4} b^2+\gamma_E\big]D_g^\sigma(\alpha,b)+{a_s\over \alpha\sigma s}\partial_\perp^2D_g^\sigma(\alpha,b)\big]
\label{interpol1}
\end{eqnarray}
Thus, the size of the correction is $\sim{1\over \alpha \sigma_as}{\partial_\perp^2D_g(\alpha,b_\perp;\sigma)\over D_g(\alpha,b_\perp;\sigma)}$.  From the model (\ref{mvodel}) we get an estimate
\beq
{1\over \alpha\sigma s}{\partial^2D_g^\sigma(\alpha,b)\over D_g^\sigma g(\alpha,b)}
\simeq~{\tigma_p\over\sigma}\Big[Q_s^2b^2+{Q_s^4\over 4}b^4+...\Big]
\eeq
Thus,  judging by the size of first correction, Sudakov evolution can be extended somewhat lower than $\tigma_p$.

Next, let us estimate the correction to the BFKL evolution.

\subsection{Correction to the BFKL evolution}
The BFKL evolution can be applied to $\sigma_a$ from the end of Sudakov region ($\sim \tigma_p$) to $\sigma_0\sim {m_n^2\over s}$ -  
the final point of the BFKL evolution. At this final point, we will use the Eq. (\ref{mvodel}) model.
The interpolation equation (\ref{interpol}) in the BFKL region can be rewritten as
\bega
&&\hspace{-1mm}
\sigma{d\over d\sigma}D_g^\sigma(\alpha,z)~=~{a_s\over\pi}\!\int\!dz’\Big[{D_g^\sigma(\alpha,z’)
\over (z-z’)_\perp^2}-{(z,z’)\over (z-z’)^2{z’}^2}D_g^\sigma(\alpha,z)\Big]
\nonumber\\
&&\hspace{22mm}
+~{a_s\over\pi}\!\int\!dz {e^{-\vigma(z-z’)^2}-1\over (z-z’)_\perp^2}D_g^\sigma(\alpha,z’)
\label{interpol2}\\
&&\hspace{-1mm}
\simeq~~{a_s\over\pi}\!\int\!dz’\Big[{D_g^\sigma(\alpha,z’)\over (z-z’)^2}
-{(z,z’)\over (z-z’)^2{z’}^2}D_g^\sigma(\alpha,z)\Big]
+{a_s\over 2\pi}\vigma^2\!\int\! dz’~{z’}^2D_g^\sigma(\alpha,z’)~+~...
\nonumber
\ega

One can estimate relative weight 
of correction in Eq. (\ref{interpol2}) by taking Eq. (\ref{mvodel}) as TMD in the RHS of Eq. (\ref{interpol2}).  One obtains 
\bega
&&\hspace{-1mm}
\sigma{d\over d\sigma}D_g^\sigma(\alpha,z)~\stackrel{\sigma\sim \sigma_0}\simeq~
-a_sD_g^\sigma(\alpha,z)\big[\ln{Q_s^2z^2\over 4}+\gamma_E\big]+8a_sCQ_s^2\Big({x_A\sigma s\over Q_s^2}\Big)^2
\nonumber\\
\ega
The relative weight of the correction is  
$\Big({x_A\sigma s\over Q_s^2}\Big)^2\sim~\Big({\sigma\over\tigma_p}{b_\perp^{-2}\over Q_s^2}\Big)^2$ so it becomes $\sim 1$ when
$\sigma_p^\star=\tigma_pQ_s^2b^2\sim\tigma_p{Q_s^2\over q_\perp^2}$.

In both cases it appears that the transition between Sudakov and BFKL evolution occurs at 
$\sigma_p^\star=\tigma_pQ_s^2b^2$ and $\sigma_t^\star=\tigma_tQ_s^2b^2$ which
are somewhat lower than $\tigma_p$ andr $\tigma_t$ for $b^{-2}\sim q_\perp^2\gg Q_s^2$.
In the next Section we will discuss possible effects due to that.

\section{Hadronic tensor for particle production by gluon fusion }
As mentioned above, if we model gluon-gluon-Higgs triangle by a local vertex, the result for Sudakov evolution of the hadronic tensor is given by Eq. (\ref{momentumresult}). 
If we stop Sudakov evolution at $\tigma_p = {4b_\perp^{-2}\over x_A s}$ and $\tigma_t = {4b_\perp^{-2}\over x_B s}$ we get
\begin{eqnarray}
&&\hspace{-1mm}
W( x_A, x_B,b_\perp)
~=~{\pi^2\over 2}Q^2\pizg_{ij}^{\tigma_p}( x_A,b_\perp)\pizg^{ij;\tigma_t}( x_B,b_\perp)
\nonumber\\
&&\hspace{2mm}
\times~~\exp~\Big\{{a_s\over 2}\Big[-\Big(\ln{Q^2b_\perp^2\over 4}+2\gamma_E\Big)^2+2\gamma_E^2+{\pi^2\over 2}\Big]\Big\}
\label{sudakovresult1}
\end{eqnarray}
On the contrary, if we extend the Sudakov evolution to $\sigma_p^\star$ and  $\sigma_t^\star$, we obtain
\begin{eqnarray}
&&\hspace{-1mm}
W( x_A, x_B,b_\perp)
~=~{\pi^2\over 2}Q^2\pizg_{ij}^{\sigma_p^\star}( x_A,b_\perp;p_A)\pizg^{ij;\sigma_t^\star}( x_B,b_\perp;p_B)
\nonumber\\
&&\hspace{2mm}
\times~~\exp~\Big\{{a_s\over 2}\Big[-\Big(\ln{Q^2b_\perp^2\over 4}+2\gamma_E\Big)^2+2\Big(\ln{Q_s^2b_\perp^2\over 4}+\gamma_E\Big)^2
+{\pi^2\over 2}\Big]\Big\}
\label{sudakovresult2}
\end{eqnarray}
To compare the numerical difference between these expressions,  let us take into account only $D_g$ part of Eq. (\ref{operde}) and use  the Eq. (\ref{mvodel}) model for gluon TMDs.
The difference is shown  in  Fig. \ref{fig:plot} below.
\begin{figure}[htb]
\begin{center}
\includegraphics[width=77mm]{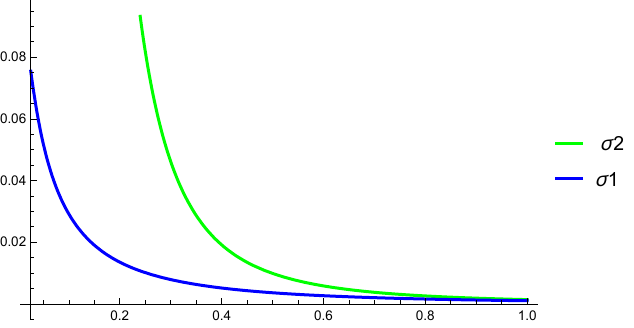}
\end{center}
\caption{The difference between Eq. (\ref{sudakovresult1}) with $\sigma1=\tigma$ and 
Eq. (\ref{sudakovresult2}) with $\sigma2=\sigma_\ast$. The $X$ axis is in GeV$^{-1}$ and $Y$ axis 
is up to an overall constant.
\label{fig:plot}}
\end{figure}
From Fig. \ref{fig:plot} we see that starting from $z\sim 0.5$ GeV$^{-1}$, corresponding to $q_\perp\simeq 2$ GeV, the difference becomes essential
so more careful analysis of interpolation equation (\ref{interpol}) is necessary.

\section{Conclusions}
The main conclusion is that the Eq. (\ref{interpol}) describes interpolation between Sudakov and BFKL evolutions of small-$x$ gluon TMDs. 
The ultimate goal would be to solve this equation and find the correct transition between Sudakov and BFKL evolutions in Eq. (\ref{momentumresult}). 
Alternatively, one may use the same logic as in hybrid factorization:  Sudakov evolution up to some matching point $\sigma$ with subsequent BFKL evolution.
With this method, the numerical solution of interpolation equation (\ref{interpol}) is sufficient since it can determine
 the optimal point 
for switching from Sudakov evoluion th the BFKL/BK one and the accuracy of this hybrid factorization.  The preliminary analysis performed in this paper shows that the matching point appears 
lower than naive point $\tigma={4b_\perp^{-2}\over x s}$, namely at $\sigma_\ast={4Q_s^2\over x s}$. As we see  from Fig. \ref{fig:plot},  starting from $q_\perp\geq 2$ GeV
the difference in estimates of $W(q)$ appears to be essential so a more careful analysis of interpolation equation (\ref{interpol}) is required. 
In conclusion let me mention that the analysis carried out in this paper can help in other cases where hybrid factorization is used, see e.g. Refs.  \cite{Iancu:2016vyg,vanHameren:2020rqt,Taels:2022tza, vanHameren:2025hyo} for recent activity.

The author is grateful to A. Vladimirov for valuable comments. This work is supported by DOE grant DE-FG02-97ER41028 
and by  Saturated Glue (SURGE) Topical Theory Collaboration. 

\section{Appendix: Interpolation equation in the Mellin representation}

In this Section we rewrite  interpolation equation (\ref{interpol}) using Mellin representation 
which solves BFKL equation (\ref{bfklevol}):
\begin{eqnarray*}
&&\hspace{0mm}
D_g^\sigma(\alpha,\nu)~=~\!\int\! dz_\perp^2(z_\perp^2)^{-\half -i\nu}D_g^\sigma(\alpha,z_\perp)
\label{mellin1}\\
&&\hspace{-0mm}
D_g^\sigma(\alpha,z_\perp)~=\int\! {d\nu\over 2\pi}(z_\perp^2)^{-\half +i\nu}D_g^\sigma(\alpha,\nu)
\label{mellin2}
\end{eqnarray*}

Substituting Eq. (\ref{mellin1}) to Eq. (\ref{evold}) one obtains
\begin{eqnarray}
&&\hspace{-1mm}
\sigma{d\over d\sigma}D_g^\sigma(\alpha,\nu)~=~{a_s\over\pi^2b^2}\int\!{d\nu'\over2\pi}\!\int\! d^2x
(x^2)^{-\half -i\nu}
\nonumber\\
&&\hspace{-1mm}
\times~\! \int\!d^2z\Big[{e^{-\alpha\sigma s(x-z)^2/4}\over (x-z)^2}(z^2)^{-\half +i\nu’}-{(x,z)\over (x-z)^2z^2}(x^2)^{-\half +i\nu’}\bigg]D_g^\sigma(\alpha,\nu’)
\end{eqnarray}
where we use notation $\vigma\equiv \alpha\sigma {s\over 4}b^2$. Regularizing divergent integrals 
${1\over (x-z)^2}\rightarrow{x^{-2\ve}\over (x-z)^2)^{1-\ve}}$ we get
\begin{eqnarray}
&&\hspace{-1mm}
{1\over\pi^2}\!\int\!d^2x (x^2)^{-\half-i\nu}\!\int\!d^2z\Big[(x^2)^{-\ve}{e^{-\alpha\sigma s(x-z)^2/4}\over [(x-z)^2]^{1-\ve}}(z^2)^{-\half+i\nu’}
-{(x,z)\over [(x-z)^2]^{1-\ve}z^2} (x^2)^{-\half +i\nu’-\ve}\Big]
\nonumber\\
&&\hspace{-1mm}
=~{1\over\pi^2}\!\int\!d^2z{e^{-\varsigma z^2}\over (z^2)^{1-\ve}}
\!\int\!d^2x(x^2)^{-\ve-\half-i\nu}[(z+x)^2]^{-\half+i\nu’}-{1\over\ve}\delta(\nu-\nu’)
\nonumber\\
&&\hspace{-1mm}
=~{\Gamma(i\nu’-i\nu)\over (\varsigma/b^2)^{i\nu’-i\nu}}
{\Gamma(i\nu-i\nu’+\ve)\Gamma\big(\half-i\nu-\ve))\Gamma\big(\half+i\nu’)\over \Gamma(1+i\nu’-i\nu-\ve)\Gamma\big(\half+i\nu+\ve))\Gamma\big(\half-i\nu’)}
-{1\over\ve}\delta(\nu-\nu’)
\end{eqnarray}
From the interpolation equation (\ref{interpol})  we obtain
\begin{eqnarray*}
&&\hspace{-1mm}
\sigma{d\over d\sigma}D_g(\alpha,\gamma;\sigma)~=~a_s
\lim_{\ve\rightarrow 0}\bigg[{\Gamma(1-\gamma-\ve)\over\Gamma(\gamma+\ve)}
\label{evoleqnu}\\
&&\hspace{-1mm}
\times\int_{\gamma+{\ve\over 2}-i\infty}^{\gamma+{\ve\over 2}+i\infty}\!{d\xi\over 2\pi i}\big({\alpha s\sigma\over 4}\big)^{\gamma-\xi}
{\Gamma(\xi-\gamma)\Gamma(\gamma-\xi+\ve)\Gamma(\xi)\over \Gamma(1+\xi-\gamma-\ve)\Gamma(1-\xi)}D_g(\alpha,\xi;\sigma)
-{1\over\ve}D_g(\alpha,\gamma;\sigma)\bigg]   
\nonumber
\end{eqnarray*}
where we introduced the notations $\gamma=\half+i\nu$ and $\xi=\half+i\nu’$.  

To take limit $\ve\rightarrow 0$ we move contour of integration to the left, taking residue at $\xi=\gamma$ which gives
$\chi(\gamma)=-2\gamma_E-\psi(\gamma)-\psi(1-\gamma)$  at $\ve\rightarrow 0$. After that, one can put $\ve=0$ in the contour integral and get
\begin{eqnarray}
&&\hspace{-1mm}
\sigma{d\over d\sigma}D_g^\sigma(\alpha,\gamma)~=~
\label{melleq}\\
&&\hspace{-1mm}
=~a_s\bigg[\chi(\gamma)D_g^\sigma(\alpha,\gamma) -{\Gamma(1-\gamma)\over\Gamma(\gamma)}
\! \int_{{1\over 4}-i\infty}^{{1\over 4}+i\infty}\!{d\xi\over 2\pi i}\big({\alpha s\sigma\over 4}\big)^{\gamma-\xi}
{\Gamma(\gamma-\xi)\Gamma(\xi)\over (\gamma-\xi)\Gamma(1-\xi)}D_g^\sigma(\alpha,\xi)\bigg]
\nonumber
\end{eqnarray}
It is convenient to introduce $\tilD_g^\sigma(\alpha,\gamma)={\Gamma(\gamma)\over\Gamma(1-\gamma)}D^\sigma(\alpha,\gamma)$ and get
\begin{eqnarray}
&&\hspace{-1mm}
\sigma{d\over d\sigma}\tilD^\sigma(\alpha,\gamma)~=~a_s
\bigg[\chi(\gamma)\tilD^\sigma(\alpha,\gamma) -
\! \int_{{1\over 4}-i\infty}^{{1\over 4}+i\infty}\!{d\xi\over 2\pi i}\big({\alpha s\sigma\over 4}\big)^{\gamma-\xi}
{\Gamma(\gamma-\xi)\over (\gamma-\xi)}\tilD^\sigma(\alpha,\xi)\bigg]
\end{eqnarray}

To get connection to Sudakov evolution,  we can move contour in Eq. (\ref{melleq}) to the right, take residue at $\xi=\gamma$ and get
\begin{eqnarray}
&&\hspace{-1mm}
\sigma{d\over d\sigma}D_g^\sigma(\alpha,\gamma)~=~a_s
\bigg[
\Big({\partial\over\partial\gamma}-\ln{\alpha s\sigma\over 4}-\gamma_E\Big)D_g^\sigma(\alpha,\gamma)
\\
&&\hspace{-1mm}
-~{\Gamma(1-\gamma)\over\Gamma(\gamma)} \int_{{3\over 4}-i\infty}^{{3\over 4}+i\infty}\!{ds\over 2\pi i}\big({\alpha s\sigma\over 4}\big)^{\gamma-\xi}
{\Gamma(\gamma-s)\Gamma(s)\over (\gamma-s)\Gamma(1-s)}D_g^\sigma(\alpha,\xi)\bigg]
\nonumber
\end{eqnarray}
The first term in this equation corresponds to Sudakov evolution (\ref{sudevol}) and the second term is the correction.

 \bibliography{bibfail1.bib}

\begin{thebibliography}{10}

\bibitem{Collins:1984kg}
John~C. Collins, Davison~E. Soper, and George~F. Sterman.
\newblock {Transverse Momentum Distribution in Drell-Yan Pair and W and Z Boson
  Production}.
\newblock {\em Nucl. Phys.}, B250:199--224, 1985.

\bibitem{Ji:2004wu}
Xiang-dong Ji, Jian-ping Ma, and Feng Yuan.
\newblock {QCD factorization for semi-inclusive deep-inelastic scattering at
  low transverse momentum}.
\newblock {\em Phys. Rev.}, D71:034005, 2005.

\bibitem{Collins:2011zzd}
John Collins.
\newblock {\em {Foundations of perturbative QCD}}.
\newblock Cambridge University Press, 2013.

\bibitem{Collins:2014jpa}
John Collins and Ted Rogers.
\newblock {Understanding the large-distance behavior of
  transverse-momentum-dependent parton densities and the Collins-Soper
  evolution kernel}.
\newblock {\em Phys. Rev.}, D91(7):074020, 2015.

\bibitem{Balitsky:2017flc}
I.~Balitsky and A.~Tarasov.
\newblock {Higher-twist corrections to gluon TMD factorization}.
\newblock {\em JHEP}, 07:095, 2017.

\bibitem{Balitsky:2017gis}
I.~Balitsky and A.~Tarasov.
\newblock {Power corrections to TMD factorization for Z-boson production}.
\newblock {\em JHEP}, 05:150, 2018.

\bibitem{Balitsky:2019ayf}
Ian Balitsky and Giovanni~A. Chirilli.
\newblock {Conformal invariance of transverse-momentum dependent parton
  distributions rapidity evolution}.
\newblock {\em Phys.\ Rev.\ D}, 100(5):051504.

\bibitem{Balitsky:2020jzt}
Ian Balitsky.
\newblock {Gauge-invariant TMD factorization for Drell-Yan hadronic tensor at
  small x}.
\newblock {\em JHEP}, 05:046, 2021.

\bibitem{Balitsky:2021fer}
Ian Balitsky.
\newblock {Drell-Yan angular lepton distributions at small x from TMD
  factorization.}
\newblock {\em JHEP}, 09:022, 2021.

\bibitem{Balitsky:2022vnb}
Ian Balitsky and Giovanni~A. Chirilli.
\newblock {Rapidity evolution of TMDs with running coupling}.
\newblock {\em Phys. Rev. D}, 106(3):034007, 2022.

\bibitem{Balitsky:2023hmh}
Ian Balitsky.
\newblock {Rapidity-only TMD factorization at one loop}.
\newblock {\em JHEP}, 03:029, 2023.

\bibitem{Fadin:1975cb}
Victor~S. Fadin, E.~A. Kuraev, and L.~N. Lipatov.
\newblock {On the Pomeranchuk Singularity in Asymptotically Free Theories}.
\newblock {\em Phys. Lett. B}, 60:50--52, 1975.

\bibitem{Balitsky:1978ic}
I.~I. Balitsky and L.~N. Lipatov.
\newblock {The Pomeranchuk Singularity in Quantum Chromodynamics}.
\newblock {\em Sov. J. Nucl. Phys.}, 28:822--829, 1978.

\bibitem{Balitsky:1995ub}
I.~Balitsky.
\newblock {Operator expansion for high-energy scattering}.
\newblock {\em Nucl. Phys. B}, 463:99--160, 1996.

\bibitem{Kovchegov:1999yj}
Yuri~V. Kovchegov.
\newblock {Small x F(2) structure function of a nucleus including multiple
  pomeron exchanges}.
\newblock {\em Phys. Rev.}, D60:034008, 1999.

\bibitem{Kovchegov:1999ua}
Yuri~V. Kovchegov.
\newblock {Unitarization of the BFKL pomeron on a nucleus}.
\newblock {\em Phys. Rev.}, D61:074018, 2000.

\bibitem{McLerran:1993ni}
Larry~D. McLerran and Raju Venugopalan.
\newblock {Computing quark and gluon distribution functions for very large
  nuclei}.
\newblock {\em Phys. Rev. D}, 49:2233--2241, 1994.

\bibitem{McLerran:1993ka}
Larry~D. McLerran and Raju Venugopalan.
\newblock {Gluon distribution functions for very large nuclei at small
  transverse momentum}.
\newblock {\em Phys. Rev. D}, 49:3352--3355, 1994.

\bibitem{Golec-Biernat:1998zce}
Krzysztof~J. Golec-Biernat and M.~Wusthoff.
\newblock {Saturation effects in deep inelastic scattering at low Q**2 and its
  implications on diffraction}.
\newblock {\em Phys. Rev. D}, 59:014017, 1998.

\bibitem{Dumitru:2005gt}
Adrian Dumitru, Arata Hayashigaki, and Jamal Jalilian-Marian.
\newblock {The Color glass condensate and hadron production in the forward
  region}.
\newblock {\em Nucl. Phys. A}, 765:464--482, 2006.

\bibitem{Chirilli:2011km}
Giovanni~A. Chirilli, Bo-Wen Xiao, and Feng Yuan.
\newblock {One-loop Factorization for Inclusive Hadron Production in $pA$
  Collisions in the Saturation Formalism}.
\newblock {\em Phys. Rev. Lett.}, 108:122301, 2012.

\bibitem{Chirilli:2012jd}
Giovanni~A. Chirilli, Bo-Wen Xiao, and Feng Yuan.
\newblock {Inclusive Hadron Productions in pA Collisions}.
\newblock {\em Phys. Rev. D}, 86:054005, 2012.

\bibitem{Mueller:2013wwa}
A.~H. Mueller, Bo-Wen Xiao, and Feng Yuan.
\newblock {Sudakov double logarithms resummation in hard processes in the
  small-x saturation formalism}.
\newblock {\em Phys. Rev. D}, 88(11):114010, 2013.

\bibitem{Balitsky:2015qba}
I.~Balitsky and A.~Tarasov.
\newblock {Rapidity evolution of gluon TMD from low to moderate x}.
\newblock {\em JHEP}, 10:017, 2015.

\bibitem{Mulders:2000sh}
P.~J. Mulders and J.~Rodrigues.
\newblock {Transverse momentum dependence in gluon distribution and
  fragmentation functions}.
\newblock {\em Phys. Rev.}, D63:094021, 2001.

\bibitem{Dominguez:2011wm}
Fabio Dominguez, Cyrille Marquet, Bo-Wen Xiao, and Feng Yuan.
\newblock {Universality of Unintegrated Gluon Distributions at small x}.
\newblock {\em Phys. Rev. D}, 83:105005, 2011.

\bibitem{Balitsky:2016dgz}
I.~Balitsky and A.~Tarasov.
\newblock {Gluon TMD in particle production from low to moderate x}.
\newblock {\em JHEP}, 06:164, 2016.

\bibitem{Boer:2016fqd}
Dani{\"e}l Boer, Piet~J. Mulders, Cristian Pisano, and Jian Zhou.
\newblock {Asymmetries in Heavy Quark Pair and Dijet Production at an EIC}.
\newblock {\em JHEP}, 08:001, 2016.

\bibitem{Iancu:2016vyg}
E.~Iancu, A.~H. Mueller, and D.~N. Triantafyllopoulos.
\newblock {CGC factorization for forward particle production in proton-nucleus
  collisions at next-to-leading order}.
\newblock {\em JHEP}, 12:041, 2016.

\bibitem{vanHameren:2020rqt}
A.~van Hameren, P.~Kotko, K.~Kutak, and S.~Sapeta.
\newblock {Sudakov effects in central-forward dijet production in high energy
  factorization}.
\newblock {\em Phys. Lett. B}, 814:136078, 2021.

\bibitem{Taels:2022tza}
Pieter Taels, Tolga Altinoluk, Guillaume Beuf, and Cyrille Marquet.
\newblock {Dijet photoproduction at low x at next-to-leading order and its
  back-to-back limit}.
\newblock {\em JHEP}, 10:184, 2022.

\bibitem{vanHameren:2025hyo}
Andreas van Hameren and Maxim Nefedov.
\newblock {Hybrid high-energy factorization and evolution at NLO from the
  high-energy limit of collinear factorization}.
\newblock {\em JHEP}, 02:160, 2025.

\end{thebibliography}
 \end{document}